\documentclass[12pt,aps,prb,preprint]{revtex4}
\usepackage{mathtext, amsmath}
\usepackage{graphicx}
\usepackage[OT1,OT2,T2A]{fontenc}
\usepackage[cp1251]{inputenc}        
\usepackage[english]{babel}
\usepackage{citehack}
\usepackage{verbatim}
\begin{document}
\title{The equation of state of the n-vector model: collective variables method.}

\author{
P.R. Kozak, M.P. Kozlovskii, Z. Usatenko}

\affiliation{Institute for Condensed Matter Physics of the
National Academy of Sciences of Ukraine, 1 Svientsitskii Str.,
79011 Lviv, Ukraine}

\newcommand{\figsize}{0.618\textwidth} 
\newcommand{\no}{\nonumber}
\newcommand{\non}{\nonumber \\}
\newcommand{\vh}{\mathbf h}
\newcommand{\vi}{\mathbf{i}}
\newcommand{\vl}{\mathbf{l}}
\newcommand{\vj}{\mathbf{j}}
\newcommand{\vk}{\mathbf{k}}
\newcommand{\vm}{\mathbf{m}}
\newcommand{\vrho}{\boldsymbol{\rho}}
\newcommand{\veta}{\boldsymbol{\eta}}
\newcommand{\rhomk}{{\rho_{-\vk}}}
\newcommand{\vS}{\boldsymbol{S}}
\newcommand{\omk}{{\omega_\vk}}
\newcommand{\ommk}{{\omega_{-\vk}}}
\newcommand{\etak}{{\eta_\vk}}
\newcommand{\etamk}{{\eta_{-\vk}}}

\newcommand{\be}{\begin{equation}}
\newcommand{\ee}{\end{equation}}
\newcommand{\bea}{\begin{eqnarray}}
\newcommand{\eea}{\end{eqnarray}}
\newcommand{\lp}{\left (}
\newcommand{\rp}{\right )}
\newcommand{\lbr}{\left \{}
\newcommand{\rbr}{\right \}}
\newcommand{\lb}{\left [}
\newcommand{\rb}{\right ]}
\newcommand{\ld}{\left .}
\newcommand{\rd}{\right .}
\newcommand{\cH}{{\cal H}}
\newcommand{\cB}{{\cal B}}
\newcommand{{\cM}}{{\cal M}}
\newcommand{\cR}{{\cal R}}

\begin{abstract}
The critical behavior of  the three-dimensional n-vector model in the presence of an external field is investigated. Mathematical description is performed with the collective variables (CV) method in the framework of the $\rho^4$ model approximation at the microscopic level without any adjustable parameters. The  recurrence relations of the renormalization group (RG) as functions of the external field and temperature were found. The analytical expression for the free energy of the system for temperatures $T>T_c$ and different n was obtained.  The equation of state of the n-vector model for general case of small and large external fields was written. The explicit form of the correspondent scaling function for different values of the order parameter was derived. The obtained results are in qualitative  agreement with the data of Monte Carlo simulations. 
\end{abstract}

\maketitle

\section{Introduction}

 The investigation of the critical behavior of the real
three-dimensional(3D) magnets is one of the most important problems
of condensed matter physics. The present work is connected with
investigation of the classical n-vector model on 3D simple cubic
lattice in the presence of an external magnetic field by the
collective variables (CV) method. Originally this method was
introduced by Bom \cite{Bom}, then used by Zubarev for systems of
charged particles \cite{Zub} and later developed for calculation of
the thermodynamic and structural characteristics of 3D systems near
the phase transition (PT)  point \cite{06}. The above mentioned
model is well-known as the classical $O(n)$-vector model or, in
field-theoretic language, as the $O(n)$-invariant nonlinear
$\sigma$-model. Depending on components of an order parameter this
model can describe a number of physical systems such as: polymers,
ferromagnets, antiferromagnets, the critical point of the
liquid-vapor transition, the Bose-condensation, phase transitions in
binary alloys etc.

The investigations of the critical properties of the O(n)-vector
models and their partial cases were carried out by various methods
such as: high- and low-temperature expansions, the field theory, the
semi-microscopic scaling field theory and Monte-Carlo simulations.
In general, much attention was devoted to investigation of the
universal characteristics of the system such as critical exponents
and relations of the critical amplitudes of the thermodynamic
functions.

The collective variables method as well as Wilson's approach
\cite{Wil} is based on use of the hypothesis of scaling invariance
and the renormalization group (RG) method for the phase transition
theory suggested by Patashynskii, Pokrovskii \cite{Pok} and Kadonoff
\cite{Kad}.

The RG method was used to obtain the equation of state of Ising
system up to the order $\epsilon^2$ by Avdeiva and Migdal \cite{Avd}
and by Bresin, Wallace, Wilson \cite{Briz}. The obtained results
were generalized for the case of the n-vector model in \cite{Briz1}.

Besides, the positive results were achieved at calculation of the
thermodynamic functions near the critical point. In Wegner's work
\cite{Weg} was obtained expression for the free energy with taking
into account the so-called `irrelevant' operators in the Wilson's
approach. Riedel and Wegner \cite{Weg1} suggested the method of
scaling fields for obtaining crossover scaling functions of the free
energy and the susceptibility. The works of Fisher and Aharony
\cite{Fish}, Nicoll and Albright \cite{Nik} and also Nelson
\cite{Nel} are dedicated to receiving of the crossover scaling
functions for $T>T_c$ in zero magnetic field near four dimensions.
In the frame of the massive field theory by Bagnuls and Bervillier
\cite{Bag} the explicit results for the correlation length, the
susceptibility and the heat capacity as functions of the temperature
in the disordered phase along the critical isochore for one-, two-
and three-component systems were obtained. The non asymptotic
behavior was described as crossover between the Wilson-Fisher's
(near the critical temperature $T_c$) and mean field's (far from
$T_c$) behaviors using three adjustable parameters. But this
crossover can not realistically describe the situation in the
system, because there are some physical restrictions of the model.
Thus, in the works of Dohm and co-workers \cite{Domb1,Domb2} the
calculation of the thermodynamical characteristics of the system
without $\epsilon$-expansion was performed in the frame of some
minimal subtraction scheme based on high-ordered perturbation theory
and Borel resummation. This minimizing scheme is related to the use
of the general relations between the heat capacity coefficients for
approximation of the temperature dependence of coefficient $u(t)$
near the fourth term in the Ginsburg-Landau Hamiltonian. This method
allows to obtain the non universal critical behavior of the
thermodynamic functions below and above the critical temperature,
such as the heat capacity and the susceptibility as functions of
$u(t)$ without any adjustable parameters. But it does not provide
possibility to analyze the dependence of the thermodynamic variables
on the microscopic parameters of the interaction potential.

Besides, the essential success was achieved in calculation of the
universal relations of the critical amplitudes. Okabe and Ohno
\cite{Ok1}, Okabe and Ideura \cite{Ok2} investigated the relations
of the critical amplitudes of the susceptibility by
high-temperature-, 1/n- and $\epsilon$-expansion up to the order
$O(\epsilon^2)$. Bresin, Le Guillon and Zinn-Justin \cite{Brez}
calculated the universal relations of the critical amplitudes for
the heat capacity, the susceptibility and the correlation length by
Wilson-Fisher's $\epsilon$-expansion.

It should be mentioned, that PT is actively investigated by
Monte-Carlo (MC) method. Thus Ferrenberg and Landau \cite{Fer1,
Fer2} found the critical temperature and critical exponents for the
Ising and classical Heisenberg models using high-resolution MC
method. Besides, the universal relations of the critical amplitudes
and the equation of state were obtained by Engels for
$O(1),\;O(2),\;O(4)$ models \cite{Eng1, Eng2, Eng4} and  by Campostrini  et al. for $O(3)$ model \cite{Camp}.

A number of new results were obtained using for the description of
PT the collective variables method. The specificity of the CV method
is successive microscopic approach and the method of integration of
the partition function by short-wave fluctuations without applying
of the perturbation theory. In the frame of this method the general
recurrence relations (RR) which correspond to the RG equations, the
critical exponents and the relation of the critical amplitudes of
Ising model were obtained.

Investigation of the $O(n)$ model allows to obtain, in the unified
form, results for the critical behavior of whole class of systems
such as: polymers in $n\to 0$ limit, the Ising model for n=1, the
XY-model for n=2, the Heisenberg model for n=3, the model with n=4
is important for quantum chromodynamics with two degenerate
light-quark flavours at finite temperature and the spherical model
in the case n$\rightarrow\infty$ which has the exact solution.

The quantity n is related to dimensionality of the order parameter
of the system. The investigation of the n-vector model was carried
out by the CV method in \cite{VakRudGol} using the
Stratanovich-Hubbard representation. The CV method was used for
investigation of properties of the pre-transition behavior and
description of the structural PT in the system with the n-component
order parameter \cite{YuhMr2}. The thermodynamical characteristics
of the n-vector model in zero magnetic field were found in
\cite{Usat, Usat1} using the CV method.

In general, the real physical systems are characterized by the
presence of the external fields. The description of systems with the
n-component order parameter in the presence of the external fields
is complicated task and needs detailed study. Thus, taking into
account the results obtained in \cite{Usat,Usat1}, we investigate
the influence of the external field on the critical behavior of the
n-vector model.


\section{The model.}

The Hamiltonian $H$ of the $n$-vector model in the presence of the
external field has the form:

\be
H=-\frac{1}{2}\sum_\vi\sum_\vj\Phi(|\vi-\vj|)\vS_\vi\vS_\vj-\mathbf
H\sum_\vi \vS_\vi,\label{h1} \ee
where $\vS_\vi=(S_\vi^{(1)},\ldots,S_\vi^{(n)})$ is the classical
n-component spin of length m localized at the N sites of
d-dimensional cubic lattice with coordinates $\vi$,
$\Phi\left(\left|\vi-\vj\right|\right)$ is the interaction
potential.

The partition function of the model (\ref{h1}) is the functional
integral over all possible orientations of the spin vector and can
be written in the form:
 $m$:
\be Z=\int\prod_\vi d \vS_\vi \delta(S_\vi-m)e^{-\beta H},\ m>0, \ee
where we take into account condition that length of the spin is $m$.
We will integrate the partition function in the space of CV. Let's
introduce the variables : \bea
\hat{\vrho}_\vk ^{c}&=& \frac{1}{\sqrt{N}}\sum_{\vi} \cos({\vk \vi}) \vS_\vi, \\
\hat{\vrho}_\vk^{s}&=&\frac{1}{\sqrt{N}}\sum_\vi \sin({\vk \vi}) \vS_\vi, \\
\hat{\vrho}_0&=&\frac{1}{\sqrt{N}}\sum_\vi \vS_\vi, \eea which  are
the n-component vectors. The CV $\vrho_\vk$ are introduced as
functional representation for the operators of the fluctuation of
spin density: \be \hat{\vrho}_\vk=\int\vrho_\vk
J(\vrho-\hat{\vrho})(d\vrho_\vk)^N. \ee In the CV representation the
partition function of the model \cite{06} is: \be Z=\int
\exp\left[\frac{1}{2}\sum_\vk \beta \Phi(k)\vrho_\vk
\vrho_{-\vk}+\mathbf{h}\vrho_0 \right]J[\rho](d\vrho_\vk)^N, \ee
where \be \mathbf{h}=\beta \mathbf{H}. \no \ee The Jacobian of
transition from the spin variables to the CV has the form: \be
J[\rho]=\int\prod_\vi d \vS_\vi \delta(S_\vi-m)
\delta(\rho_0-\hat\rho_0)\acute{\prod_\vk}\delta(\rho^c_\vk-\hat\rho^c_\vk)\delta(\rho^s_\vk-\hat\rho^s_\vk).
\ee
 Calculation of the partition function is performed in the
general framework of \cite{Usat, Usat1}. The main idea is that the
phase space is divided on the intervals (layers) to depend on value
$\vk$ and the interaction potential is averaged on each of this
intervals. The Fourier transform of the interaction potential is
replaced by the following approximation   \cite{Juh,Usat,Usat1}:
\bea
\Phi(k) = \lbr \begin{array}{ll} \Phi(0) (1-2b^2k^2), & k\in\cB' \\
\bar\Phi = const, & k\in \cB/\cB'.
\end{array}\rd\nonumber \label{Fi} \eea
We assume $\bar\Phi=0$. Such cutting of the potential do not affect
the general picture of the critical behavior but is appreciable when
we want to estimate the critical temperature. We use the method
suggested in \cite{21} for integration of the partition function in
the case of presence of the external field. It should be mentioned
that in our work we use the quartic measure density that allows us
to describe the PT on qualitative good level \cite{20}
 Integration by l layers
gives: \be Z=2^{\frac{n}{2}(N_{l+1}-1)}Q_0Q_1\ldots
Q_lQ^{N_{l+1}}(P_l)Z_{l+1}, \ee where $Q_l$ is the partial partition
function of l's layer.

\bea Q_0&=&Q^{N'}(u)Q^{N'}(d_0),\non
Q^{N'}(u)&=&J'[0]\exp(u'_0N'),\non
Q_l&=&Q^{N_l}(P_{l-1})Q^{N_l}(d_l),\non
Q(d_l)&=&(2\pi)^{\frac{n}{2}}\left(\frac{3}{a_4^{(l)}}\right)^{\frac{n}{4}}U\left(\frac{n-1}{2},x_l\right)\exp\left({\frac{x_l^2}{4}}\right),
\non
Q(P_{l})&=&(2\pi)^{-\frac{n}{2}}\left[s^d\frac{n+2}{3}\frac{a_4^{(l)}}{\varphi(x_{l})}\right]^{\frac{n}{4}}U\left(\frac{n-1}{2},y_{l}\right)\exp\left({\frac{y_{l}^2}{4}}\right).
 \label{stat} \eea
The function $\varphi(x_{l})$ is defined in the appendix. The non
integrated part of $Z$ has the form: \bea
Z_{l+1}&=&\int(d\vrho_\vk)^{N_{l+1}}\exp\left\{ \sqrt{N} \vh
\vrho_0-\frac{1}{2}\sum_{\vk<\cB_{l+1}}d^{\:(l+1)}(k)\vrho_\vk\vrho_{-\vk}-\right.
\non
&-&\left.\frac{a_4^{(l+1)}}{4!N_{l+1}}\sum_{\vk_1\ldots\vk_4<\cB_{l+1}}
\vrho_{\vk_1}\ldots\vrho_{\vk_4}\delta_{\vk_1\ldots \vk_4}\right\}.
\eea

The presence of the external field results in appearance of the
linear term in the exponent. Let assume that the external field is
oriented along one of the coordinate axes (e. g. x axe), we receive:
\bea Z_{l+1}&=&\int(d\vrho_\vk)^{N_{l+1}}\exp\left\{
\sqrt{N_{l+1}}a_1^{(l+1)}
\rho_0^{(1)}-\frac{1}{2}\sum_{\vk<\cB_{l+1}}d^{\:(l+1)}(k)\vrho_\vk\vrho_{-k}-\right.
\non
&-&\left.\frac{a_4^{(l+1)}}{4!N_{l+1}}\sum_{\vk_1\ldots\vk_4<\cB_{l+1}}
\vrho_{\vk_1}\ldots\vrho_{\vk_4}\delta_{\vk_1\ldots \vk_4}\right\}.
\eea For the coefficients near different powers of $\vrho_\vk$ we
have the following recurrence relations: \bea
a_1^{(l+1)}&=&a_1^{(l)}s^\frac{d}{2},\non
a_2^{(l+1)}&=&a_2^{(l)}+d^{\:(l)}(\cB_{l+1},\cB_l)M(x_l), \non
a_4^{(l+1)}&=&a_4^{(l)}s^{-d}E(x_l), \label{nrr} \eea where \bea
M(x_l)&=&N(x_l)-1, \;
N(x_l)=\frac{y_lU_n(y_l)}{x_lU_n(x_l)},\;E(x_l)=s^{2d}\frac{\varphi(y_l)}{\varphi(x_l)},
\non x_l&=&\sqrt{\frac{3}{a^{(l)}_4}}d^{\:(l)}(\cB_{l+1},\cB), \;
y_l=s^{\frac{d}{2}}U_n(x_l)\sqrt{\frac{n+2}{\varphi(x_l)}}, \non
N_l&=&N's^{-dl}, \; N'=Ns_0^{-d}. \eea For convenience the following
designation are introduced: \bea
d^{\:(l)}(\cB_{l+1},\cB)=d^{\:(l)}(0)+qs^{-2l},\;q=\beta\Phi(0)\bar{q},
\non a_1^{(l)}=s^{-l}\omega_l,\;d^{\:(l)}(0)=s^{-2l}r_l, \;
a^{(l)}_4=s^{-4l}u_l. \eea Thus, the recurrence relations
(\ref{nrr}) can be written  in the form: \bea
\omega_{l+1}&=&s^{\frac{d+2}{2}}\omega_l, \non r_{l+1}&=&s^2
[(r_l+q)N(x_l)-q], \non u_{l+1}&=&s^{4-d}u_lE(x_l). \label{rr} \eea
The initial values of $\omega_l, r_l, u_l$ are (for $l=0$): \bea
\omega_0=s_0^\frac{d}{2}h', \; r_0=a_2-\beta\Phi(0), \; u_0=a_4.\eea
In that way we passed to the parametric space of the RG
transformation. The phase transition point is represented by the
fixed point with coordinates: \bea \omega^*=0, \;
r^*=-f_n\beta\Phi(0), \; u^*=\phi_n[\beta\Phi(0)]^2,\eea where \bea
f_n=\bar q\frac{s^2[N(x^*)-1]}{s^2N(x^*)-1}, \; \phi_n=\bar
q^2\frac{3}{x^{*2}} \left[ \frac{1-s^{-2}}{N(x^*)-s^{-2}}\right]^2.
\eea Here $x^*$ is the solution of equation \cite{Usat}: \bea
s^{4+d}\varphi(y^*)=\varphi(x^*). \eea So, when $\tau=0$, $h=0$ and
$l\rightarrow\infty$ the system is in the fixed point. It is obvious
that near the critical point when $\tau\rightarrow 0, \;
h\rightarrow 0$ and for large l in the parametric space the system
will be near the fixed point. This case is called the critical
regime (CR). In the CR the recurrence relations may be expanded by
the deviation from the fixed point \bea
\left( \begin{array}{ll} \omega_{l+1}-\omega^* \\
                                                r_{l+1}-r^* \\
                                                u_{l+1}-u^*
                                                \end{array} \right ) = \cR \left( \begin{array}{ll} \omega_l-\omega^* \\
                                                r_l-r^* \\
                                                u_l-u^*
                                                \end{array} \right
                                                ).
\eea The elements of matrix $\cR$ in linear by $(x_l-x^*)$
approximation have the form: \bea R_{11}&=&s^{\frac{d+2}{2}}, \;
R_{12}=R_{21}=R_{13}=R_{31}, \non R_{22}&=&\sqrt3 s^2\mu_1, \;
R_{23}= \frac{s^2}{2\sqrt{u^*}}(\mu_0-\mu_1x^*), \non
R_{32}&=&\sqrt{3u^*}s^{4-d}\omega_1, \;
R_{33}=s^{4-d}\left(\omega_0-\frac{\omega_1x^*}{2}\right), \eea
where the following designations were introduced: \bea \mu_1&=&\mu_0
\left(a_1-\frac{q_1}{2}\right), \;
\mu_0=\sqrt{\frac{n+2}{3\varphi(x^*)}}s^{\frac{d}{2}}U_n(y^*), \non
a_1&=&\tilde{P_1}y^*r_1, \; r_1=\partial_1-\frac{q_1}{2}, \non
\tilde{P}_m&=&\frac{1}{U_n(y^*)} \left[
\frac{d^mU_n(y_l)}{dy_l^m}\right]_{y^*}, \non
\omega_1&=&\omega_0(b_1-q_1),\;\omega_0=s^{2d}\frac{\varphi(y^*)}{\varphi(x^*)},
 \non
b_1&=&\tilde{Q}_1y^*r_1, \;
\tilde{Q}_m=\frac{1}{\varphi(y^*)}\left[\frac{d^m\varphi(y_l)}{dy_l^m}\right]_{y^*},
\non R_{23}^{(0)}&=&R_{23}\sqrt{u^*}, \;
R_{32}^{(0)}=\frac{R_{32}}{\sqrt{u^*}}. \eea Action of matrix $\cR$
allows to receive the coefficients of the partition function of the
next layer. In order to receive the coefficient of l's layer we need
to act by $\cR$ l times on the coefficients of zero layer.
 \bea
\left( \begin{array}{ll} \omega_{l}-\omega^* \\
                                                r_{l}-r^* \\
                                                u_{l}-u^*
                                                \end{array} \right ) = \cR^{\:l} \left( \begin{array}{ll} \omega_0-\omega^* \\
                                                r_0-r^* \\
                                                u_0-u^*
                                                \end{array} \right
                                                ).
\eea It is easy to obtain the form of the matrix $\cR^{\:l}$ if
reduces the matrix $\cR$ to the diagonal form. In order to do it, we
must pass to the base from the eigenvectors of $\cR$. The
eigenvalues of $\cR$ are universal quantities: \bea E_1=R_{11}, \;
E_{2,3}=\frac{1}{2} \left[
R_{22}+R_{33}\pm\sqrt{(R_{22}-R_{33})^2+4R_{23}R_{32}}\:\right].
\eea The eigenvectors have the form: \bea
\omega_1=\left(\begin{array}{ll}1\\0 \\ 0 \end{array}\right), \;
\omega_2=\left(\begin{array}{ll}0\\1 \\ R_1 \end{array}\right), \;
\omega_3=\left(\begin{array}{ll}0\\R \\ 1 \end{array}\right). \eea
The inverse vectors are written as: \bea
v_1&=&(1\:0\:0\:),\;v_2=\frac{1}{D}(0\:1\:-R),
\;v_3=\frac{1}{D}(0\:-R_1\:1),\non
R&=&\frac{R_{23}}{E_3-R_{22}},\;R_1=\frac{E_2-R_{22}}{R_{23}}. \eea
The determinant of the inverse matrix is \bea
D=\frac{E_3-E_2}{E_3-R_{22}}. \eea After expanding the coefficients
in (\ref{rr}) by eigenvectors we obtain: \bea
\omega_l&=&s_0^{\frac{d}{2}}hE^{\:l}_1,\non
r_l&=&r^*+c_1E^{\:l}_2+c_2RE^{\:l}_3,\non
u_l&=&u^*+c_1R_1E^{\:l}_2+c_2E^{\:l}_3. \eea The coefficients
$c_1,\:c_2$ can be found from the initial conditions at l=0. \bea
c_1&=&\frac{1}{D}[r_0-r^*-R(u_0-u^*)],\non
c_2&=&\frac{1}{D}[(u_0-u^*)-R_1(r_0-r^*)]. \eea

The eigenvalue $E_2>1$ is responsible for the deviation from the
fixed point, and $E_3<1$  is much more smaller than $E_2$ and we can
neglect it. This approximation neglects the confluent corrections.
Taking into account that at the phase transition point $r_l=r^*$, we
obtain the equation for the critical temperature in the form
\cite{Usat}: \be
[\beta_c\Phi(0)]^2(1-f_n-R^*\sqrt{\phi_n})-a_2\beta_c\Phi(0)+a_4R^*/\sqrt{\phi_n}=0,\ee
\be R^*=R\sqrt{u^*}. \ee This equation allows to write the solutions
of the recurrence relations as function of the temperature and the
external field for the critical regime: \bea
\omega_l&=&s_0^{\frac{d}{2}}h'E^l_1,\non
r_l&=&\beta\Phi(0)\left(-f_n+c_{1T}\tau
E_2^l+R^*c_{2T}E^l_3/\sqrt{\phi_n}\right),\non
u_l&=&[\beta\Phi(0)]^2\left(\phi_n+c_{1T}\tau\sqrt{\phi_n}R^*_1E^l_2+c_{2T}E^l_3\right).
\label{RS} \eea

The obtained coefficients are exponential functions of l. For small
l their values are small and then increase rapidly. For big l, the
value of $r_l$ (the coefficient near the square term in the
partition function) is bigger than $u_l$ (the coefficient near the
quartic term in the partition function). Thus, we can integrate the
partition function in this region using the Gaussian approximation.
But in the CR we must use the distributions of fluctuations higher
than Gaussian's one and take into account the quartic term in the
partition function.
 So let's find the number l
after what we can pass from accounting quartic to accounting only
quadratic terms. We call this number as the exit point from the CR.
In zero magnetic field this point was studied in \cite{Juh}. Let
designate it as $m_\tau$ and find it from the condition of deviation
from the fixed point: \be r_{m_\tau+1}-r^*=-\delta r^*,\;
\delta=\frac{\tau}{|\tau|} \ee As a result we obtain: \be
m_\tau=-\frac{\ln{|\tilde\tau|}}{\ln{E_2}}-1, \label{m_tau} \ee
where \be \tilde{\tau}=\tau\frac{c_{1k}}{f_n} \ee is the
renormalized reduced temperature. In the presence of the external
field the exit point from the CR is found from the condition: \be
\omega_{n_h+1}-\omega^*=h_0. \ee From this we found \cite{20}: \be
n_h=-\frac{\ln{\tilde h}}{\ln{E_1}}-1,\; \tilde
h=s_0^{\frac{d}{2}}\frac{h'}{h_0}. \label{n_h} \ee The quantity
$h_0$ is found from the normalization condition for scaling
function.

If there are both the temperature and the field the exit point
depends on relation between $\tau$ and $h$. Some boundary
temperature field $h_c$ which divides the values of fields on strong
and weak was found in \cite{Prytula,20}. The condition of this
division on strong and week fields has the form: \be m_\tau=n_h. \ee
After substituting equations for the exit points from the CR we
obtain: \be h_c=|\tilde\tau|^{p_0}, \ee where the critical exponent
$p_0$ has the form: \be
p_0=\frac{\ln{E_1}}{\ln{E_2}}=\frac{\nu}{\mu}, \; \mu=\frac{2}{d+2}.
\ee The critical exponent of the correlation length $\nu$ for $h=0$
is \be \nu=\frac{\ln{s}}{\ln{E_2}}. \ee The critical exponent $\mu$
describes the dependence of the correlation length from the field
for $T=T_c$. Therefore we can rewrite (\ref{m_tau}) in the form \be
m_\tau=-\frac{\ln{ h_c}}{\ln{E_1}}-1. \label{m_tau1} \ee In general
case deviations by the temperature and by the field from the fixed
point should be united. This united point $n_p$ can be found from
the equation \cite{Prytula}: \be
\left(s_0^{d/2}h'E_1^{n_p+1}\right)^2+\left(c_{1T}\tau\beta\Phi(0)E_2^{n_p+1}\right)^2=r^{*2}.
\ee

The value of $n_p$ can be found numerically. But numerical solution
does not allow analytically to take into account influence of the
temperature and the field on the critical behavior. Thus, in
\cite{21} formula for united exit point from the CR was introduced.
This form includes the temperature and the field and allows in the
limit of zero field or zero temperature  pass to the Eq.(\ref{n_h})
or Eq.(\ref{m_tau1}) \be n_p=-\frac{\ln{(\tilde
h^2+h_c^2)}}{2\ln{E_1}}-1. \ee Information about the exit point from
the CR allows us to divide the integration of the partition function
on two stages: integration by quartic distribution in the CR, where
values of $r_l$ and $u_l$ are commensurable and integration by
Gaussian distribution (the Gaussian regime) for the phase space
layers with $l>n_p$. But deviation from the fixed point occurs not
so sharp to exactly distinguish the critical and the Gaussian
regimes. Therefore we should take into account the transition regime
(TR) in which $r_l$ exceeds $u_l$ but we still can not use Gaussian
distribution. In order to integrate the partition function in the TR
one must use the quartic measure density. Fortunately, there is only
one layer in the phase space for which $r_l$ and $u_l$ behaves like
described for the TR. The number of this layer is next after the CR
$n_p+1$, so from the layer with $n_p+2$ the Gaussian region is
present. Thus the partition function has the form: \be
Z=\underbrace{Q_0Q_1\ldots
Q_{n_p}}_{CR}\underbrace{Q_{n_p+1}}_{TR}\underbrace{2^{\frac{n}{2}(N_{n_p+2}-1)}Q^{N_{n_p+2}}(P_{n_p+1})Z_{n_p+2}}_{LGR},
\label{Z_part}\ee where CR designates the critical region, TR is the
transition region and LGR is the limiting Gaussian region.

The coefficients near the CV change their behavior for $l>n_p$. It
simplifies calculations. Let's take $l=n_p+1$ in (\ref{stat}). Then
in order to integrate $Z_{n_p+2}$ is important to know below or
above $T_c$ is the system. The coefficients \bea
r_{n_p+2}&=&\beta\Phi(0)f_n(-1+E_2H_c),\non
u_{n_p+2}&=&[\beta\Phi(0)]^2\phi_n(1+\Phi E_2H_c),\non
\Phi&=&f_n/\sqrt \phi_n R^*_1,\  H_c=\tilde \tau E_2^{n_p+1} \eea
depend on the temperate and the field. The values of $u_{n_p+2}$ are
always positive that provide convergence of (\ref{Z_part}). The
coefficient $r_{n_p+2}$ is positive and exceeds $u_{n_p+2}$ for
large $\tau$ ($h_c>>\tilde h$). In this case the partition function
can be integrated in Gaussian approximation. But for small $\tau$
($h_c<<\tilde h$), $r_{n_p+2}$ decreases and becomes negative. In
this case the Gaussian approximation is useless. This problem can be
solved by introduction the substitution: \be
\rho^\alpha_\vk=\eta^\alpha_\vk+\sqrt N \sigma^\alpha_+\delta_\vk,
\; \vec{\sigma}_+=(\sigma_+,\ldots,0). \ee Then $Z_{n_p+2}$ changes
to: \bea Z_{n_p+2}&=&e^{NE_0(\sigma_+)}\int
(d\eta)^{N_{n_p+2}}\exp\left[-\frac{1}{2}\sum_{\vk\in\cB_{n_p+2}}d^{(n_p+2)}(k)\veta_\vk\veta_{-\vk}-\right.\non
&-&\frac{a_4^{(n_p+2)}}{12}s_0^3s^{3(n_p+2)}\sigma^2\left(\sum_{\vk\in\cB_{n_p+2}}\veta_\vk\veta_{-\vk}+2\sum_{\vk\in\cB_{n_p+2}}\eta^{(1)}_\vk\eta^{(1)}_{-\vk}\right)-\non
&-&\frac{a_4^{(n_p+2)}}{6\sqrt{N_{n_p+2}}}s_0^{\frac{3}{2}}s^{\frac{3}{2}(n_p+2)}\sigma_+\sum_{\vk_1\ldots\vk_3\in\cB_{n_p+2}}\eta_{\vk_1}^{(1)}\veta_{\vk_2}\veta_{\vk_3}\delta_{\vk_1+\vk_2+\vk_3}-\non
&-&\left.\frac{a_4^{(n_p+2)}}{24N_{n_p+2}}\sum_{\vk_1\ldots\vk_4\in\cB_{n_p+2}}\veta_{\vk_1}\dots\veta_{\vk_4}\delta_{\vk_1+\dots+\vk_4}\right]
\label{Z_{n_p+2}}, \eea where \be
E_0(\sigma_+)=h'\sigma_+-\frac{1}{2}d^{n_p+2}(0)\sigma_+^2-\frac{a_4}{24}s_0^3s^{3(n_p+2)}\sigma_+^4.
\ee The shift $\sigma_+$ is found from the condition: \be
\frac{\partial E_0(\sigma_+)}{\partial \sigma_+}=0. \ee This
condition causes the coefficient near the first power of
$\eta^{(1)}_0$ equals to zero and results to a cubic equation for
$\sigma_+$. Solution of this equation can be found in the form \be
\sigma_+=\sigma_0s^{(n_p+2)/2}. \ee Finally one obtains: \be
\sigma_0^3+p\sigma_0+q=0,\label{sigma} \ee where the following
designation were introduced: \be
p=\frac{6r_{n_p+2}}{u_{n_p+2}s_0^3},\;
q=-\frac{6h_0s^\frac{5}{2}}{u_{n_p+2}s_0^{\frac{9}{2}}}\frac{\tilde
h}{\sqrt{\tilde h^2+h_c^2}}. \ee Solutions of a cubic equation
depend on the sign of the discriminant \be
Q=\left(\frac{p}{3}\right)^3+\left(\frac{q}{2}\right)^2. \ee For
$T>T_c$ the value of Q is always positive. So the Eq.(\ref{sigma})
has one real and two complex roots. We select the real one: \bea
\sigma_0&=&A+B,\non A&=&\left(-\frac{q}{2}+\sqrt
Q\right)^\frac{1}{3}, \; B=\left(-\frac{q}{2}-\sqrt
Q\right)^\frac{1}{3}. \eea The integral (\ref{Z_{n_p+2}}) is
calculated in Gaussian approximation. This integration terminates
the calculation of the partition function: \bea
Z_{n_p+2}&=&e^{NE_0(\sigma_+)}\left(\frac{\pi}{2}\right)^{n(N_{n_p+2})}\sqrt{\frac{\pi}{d_1(0)}}\sqrt{\frac{\pi}{d_2(0)}}^{n-1}\times\non
&\times&\prod'_{\vk>0}\frac{1}{d_1(k)}\left(\frac{1}{d_2(k)}\right)^{(n-1)},
\eea where: \bea d_i(k)&=&r^{(i)}_R+\beta\Phi(0)b^2k^2,\non
r^{(1)}_R&=&s^{-2(n_p+2)}\frac{\tilde r^{(1)}_R}{2},\non
r^{(2)}_R&=&s^{-2(n_p+2)}\frac{\tilde r^{(2)}_R}{2},\non \tilde
r^{(1)}_R&=&r_{n_p+2}+\frac{1}{2}s_0^3\sigma_0^2u_{n_p+2},\non
\tilde r^{(2)}_R&=&r_{n_p+2}+\frac{1}{6}s_0^3\sigma_0^2u_{n_p+2}.
\eea Thus, dividing the phase space of the CV on layers depending on
the values of the wave vector, we distinguished two main states of
the system to refer to the critical behavior. The critical region
which corresponds to short-wave fluctuations and the Gaussian region
which corresponds to long-wave fluctuations. Short-wave fluctuations
are connected with the microscopic parameters of the model and
long-wave fluctuations determines the critical behavior. In contrast
to earlier approaches we pay equal attention on both `irrelevant
short-wave' and long-wave fluctuations.
\section{The free energy}
After calculation of the partition function we can find the free
energy of the system. \be F=-kT\ln Z. \ee Described above structure
of the partition function allows to present the free energy as a sum
of terms which corresponds to different regimes of fluctuations: \be
F=F_0+F_{CR}+F_{TR}+F_{LGR}. \ee Here\be
F_0=-kTN\ln\left[\frac{(2\pi)^\frac{n}{2}m^{n-1}}{\Gamma(n/2)}\right]
\ee is the free energy of noninteracting spins, \be
F_{CR}=-kT\sum^{n_p}_{l=0}\ln Q_{l}\label{F_{CR}} \ee is the energy
to refer to the critical region, \be F_{TR}=-kTQ_{n_p+1} \ee is the
contribution which corresponds to the transition region and,
respectively \be F_{LGR}=-kT\ln \left[
2^{\frac{n}{2}(N_{n_p+1}-N_{n_p})}Q^{N_{n_p+2}}(P_{n_p+1})Z_{n_p+2}\right]
\ee is the contribution which comes from the Gaussian region.

Eq.(\ref{F_{CR}}) depends on the exit point from the CR. In order to
know this dependence explicitly we need to sum up elements $Q_l$.
For this purpose let's extract dependence on index l in $Q_l$. The
quantity $Q_l$ is a function of $y_l$. The variable  $y$ is bigger 1
($y_l>>1$) for any temperature. Thus, we can use the expansion for
Weber's parabolic cylinder function $U(a,x)$ in $Q_l$ by inverse
powers of $y_l$. Using appropriate expansions we have: \be
F_{CR}=-kTN'f_{CR}^{0}-kT\sum^{n_p}_{l=1}N_lf_l, \ee where \bea
f_l&=&\ln
U\left(\frac{n-1}{2},x_l\right)+\frac{x_l^2}{4}+\frac{n}{2}\ln
y_{l-1}+\frac{n}{4y_{l-1}^2}(2n+7),\non
f_{CR}^{\;0}&=&u'_0+\frac{x_0^2}{4}+\frac{3u^{'2}_2}{4u'_4}+\ln
U\left(\frac{n-1}{2},z'\right)+\non &+&\ln
U\left(\frac{n-1}{2},x_0\right)+\frac{n}{4}\left[\ln
\left(\frac{3}{u'_4}\right)+\ln\left(\frac{3}{a_4}\right)\right].
\eea To extract the dependence on $l$ from $f_l$ we expand it by
powers of $x_l-x^*$ and substitute in obtained formula the solutions
of the recurrence relations. After summing up we obtain: \be
F_{CR}=-kTN'\left[\gamma
'_{01}+\gamma_1\tau+\gamma_2\tau^2-\gamma'(\tilde
h^2+h_c^2)^{\frac{3}{5}}\right], \ee where \bea
\gamma'&=&\bar\gamma_1+\bar\gamma_2H_c+\bar\gamma_3H_c^2,\non
\bar\gamma_1&=&\frac{f^*_{cr}}{1-s^{-3}},\non
\bar\gamma_2&=&\frac{f_nd_1\delta^2}{1-s^{-3}E_2},\non
\bar\gamma_3&=&\frac{f_n^2d_3\delta^4}{1-s^{-3}E_2^2}. \eea For the
free energy of the TR we obtain: \bea F_{TR}&=&-kTN'f_{n_p+1}(\tilde
h^2+h_c^2)^\frac{3}{5},\non f_{n_p+1}&=&\ln
U\left(\frac{n-1}{2},x_{n_p+1}\right)+\frac{x_{n_p+1}^2}{4}+\non
&+&\frac{n}{2}\ln y_{n_p}+\frac{n}{4y_{n_p}^2}(2n+7). \eea For the
GR we have: \bea \ln
Z_{n_p+2}&=&NE_0(\sigma_+)+\frac{n}{2}(N_{n_p+2}-1)\ln
\frac{\pi}{2}+\frac{1}{2}\ln \frac{\pi}{d_1(0)}+\non
&+&\frac{n-1}{2}\ln\frac{\pi}{d_2(0)}+\sum'_{k>0}\ln
\frac{1}{d_1(k)}+(n-1)\sum'_{k>0} \ln\frac{1}{d_2(k)}. \eea In order
to sum up by $k$ we change the summation by integration, that
results: \bea
F_{LGR}&=&F_0^{(+)}-kTN_{n_p+2}\left\{n\left[-\frac{1}{2}\ln2+\ln{s}-\frac{1}{4}\ln3+\frac{1}{4}\ln{u_{n_p+1}}-\right.\right.\non
&-&\left.\frac{1}{2}\ln{U(x_{n_p+1})}-\frac{n+2}{8y_{n_p+1}^2}\right]-\non
&-&\left. \frac{1}{2}\left[\ln\tilde
r^{(1)}_R+f'_{G_1}+(n-1)\left(\ln\tilde
r^{(2)}_R+f'_{G_2}\right)\right]\right\}, \eea where  \bea
f'_{G_i}&=&\ln( a_i^2+1)-\frac{2}{3}+\frac{2}{a_i^2}-\frac{2}{
a_i^3}\arctan a_i,\non
 a_i&=&\frac{\pi b}{c_0}\sqrt{\frac{\beta \Phi(0)}{\tilde r_R^{(i)}}}.
\eea For the convenience the following designations were introduced:
\be F_{LGR}=F_0^{(+)}+F_G, \ee where \bea
F_0^{(+)}&=&-kTNE_0(\sigma_+),\non F_G&=&-kTN_{n_p+2}f_G,\non
f_G&=&n\left[-\frac{1}{2}\ln2+\ln{s}-\frac{1}{4}\ln3+\frac{1}{4}\ln{u_{n_p+1}}-\right.\non
&-&\left.\frac{1}{2}\ln{U(x_{n_p+1})}
-\frac{n+2}{8y_{n_p+1}^2}\right]-\non &-& \frac{1}{2}\left[\ln\tilde
r^{(1)}_R+f'_{G_1}+(n-1)\left(\ln\tilde
r^{(2)}_R+f'_{G_2}\right)\right]. \eea Finally, we obtain the free
energy which is dependent on the temperature and the field: \bea
F&=&-kTN\left\{\ln
\left[\frac{(2\pi)^\frac{n}{2}m^{n-1}}{\Gamma(\frac{n}{2})}\right]-\frac{1}{s_0^3}(\gamma'_{01}+\gamma_1\tau+\gamma_2\tau^2)-\right.
\non &-&\left.e_0h'\left(\tilde
h^2+h_c^2\right)^{\frac{1}{10}}-\left(\gamma^{+}_s-e_2\right)\left(\tilde
h^2+h_c^2\right)^{\frac{3}{5}}\right\}, \eea where \bea
e_0&=&\frac{\sigma_0}{\sqrt s},\non
e_2&=&\frac{\sigma_0^2}{2s^3}\left(r_{n_p+2}+\frac{1}{12}u_{n_p+2}s_0^3\sigma_0^2\right)\non
\gamma_s^+&=&\frac{1}{s_0^3}\left(f_{n_p+1}-\gamma
'+\frac{f_G}{s^3}\right). \eea

\section{The order parameter.}
From the formula for the free energy we obtain the order parameter
of the system by direct differentiation by the field: \be
M=-\frac{1}{N}\left(\frac{dF}{dh}\right)_T \no. \ee The structure of
the free energy allows to separately differentiate parts connected
with different fluctuation processes. The result reduces to the
form: \be M=\sigma_{00}^{+}(\tilde h^2+h_c^2)^{\frac{1}{10}}.
\label{M} \ee The quantity $\sigma_{00}^+$ depends on the variable
$\alpha$ which represents the ratio between the field and the
temperature: \bea
\sigma_{00}^+&=&e_0\left(1+\frac{1}{5}\frac{\alpha^2}{1+\alpha^2}\right)+e_{00}\frac{\alpha}{\sqrt{1+\alpha^2}}+e_{02},\non
\alpha &=&\tilde{h}/h_c. \eea

Below there are equations for the coefficients that depend on
$\alpha$, n and other parameters of the model. The method of
calculation of the order parameter for the n-vector model is
analogous to the method for the Ising model \cite{21}. For
convenience we use the same designations, but in n-vector model
appears dependence on n: \bea e_0&=&\frac{\sigma_0}{\sqrt s}, \non
e_{00}&=&\frac{6s_0^{3/2}}{5h_0}(\gamma_s^+-e_2),\non
e_{02}&=&\frac{s_0^{3/2}}{h_0}\left(f_{\gamma_1}+\sigma_0^2q_s\left[1+\frac{1}{12}q_l\sigma_0^2\right]\right),
\no \eea \bea f_{\gamma_1}&=&
\frac{1}{s_0^3}\left(\gamma_p+f_p+\frac{f_{gv}}{s^3}\right),\non
f_{gv}&=&-\frac{n}{4}\frac{\Phi H_{cd}}{1+\Phi H_{cd}}+
\left[\frac{n(n+2)}{4}\frac{r_{p+1}}{y_{n_p+1}^2}-\frac{nU'(x_{n_p+1})}{2U(x_{n_p+1})}\right]g_{p+1}-
\non &-&\frac{1}{2}\left[\frac{\tilde g_R^{(1)}}{\tilde
r_R^{(1)}}+a_g^{(1)}g_a^{(1)}+(n-1)\left(\frac{\tilde
g_R^{(2)}}{\tilde r_R^{(2)}}+a_g^{(2)}g_a^{(2)}\right)\right], \non
q_s&=&\frac{\beta\Phi(0)}{2s^3}H_{cd}E_2f_n, \
q_l=\beta\Phi(0)\Phi\phi_n \frac{s_0^3}{f_n}, \non
f_p&=&\frac{n}{2}\left[r_pg_p\left(1-\frac{2n+7}{y_{n_p}^2}\right)-g_{p+1}U_n(x_{n_p+1})\right],\non
\gamma_p&=&H_{cd}(\bar \gamma_2+2 \bar \gamma_3H_c). \eea

\begin{figure}
    \centering
        \includegraphics[width=\figsize]{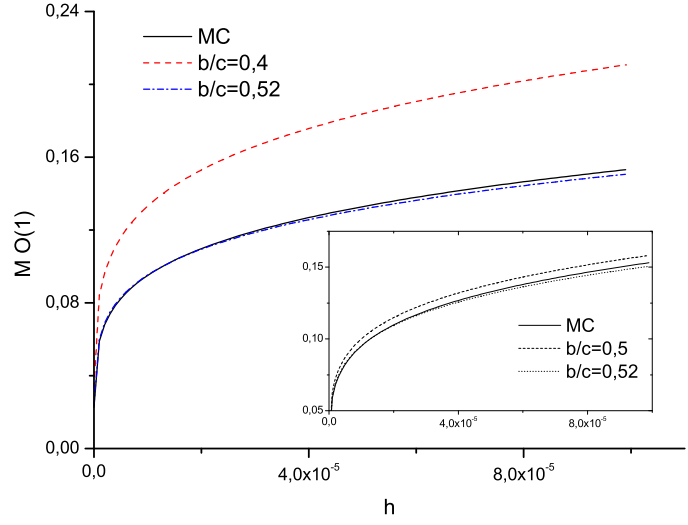}
    \caption{The dependence of the order parameter on the field for $\tau=0$ and n=1}
    \label{M1}
\end{figure}

\begin{figure}
    \centering
        \includegraphics[width=\figsize]{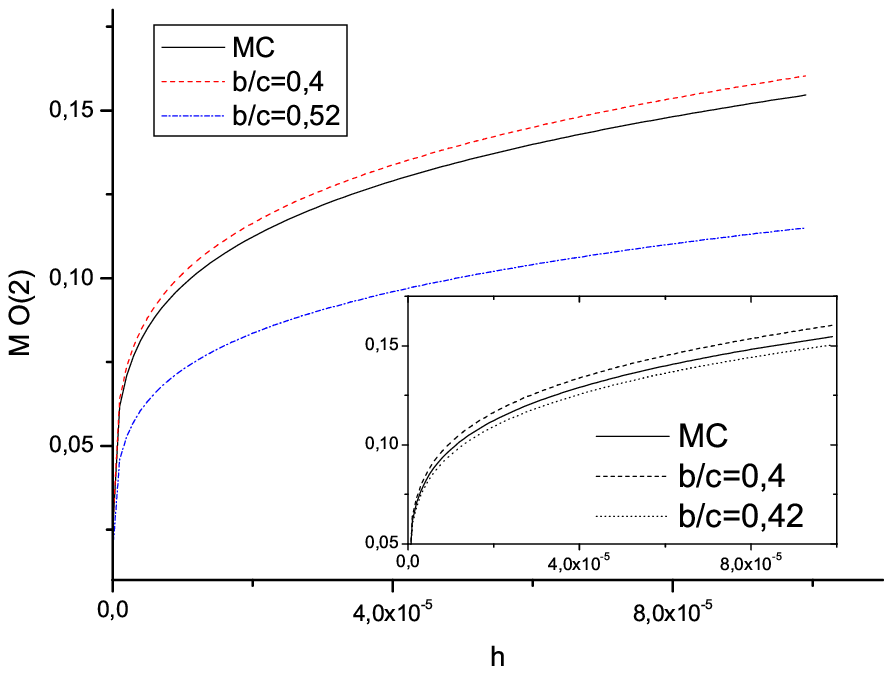}
    \caption{The dependence of the order parameter on the field for $\tau=0$ and n=2}
    \label{M2}
\end{figure}

\begin{figure}%
	\centering
	\includegraphics[width=\figsize]{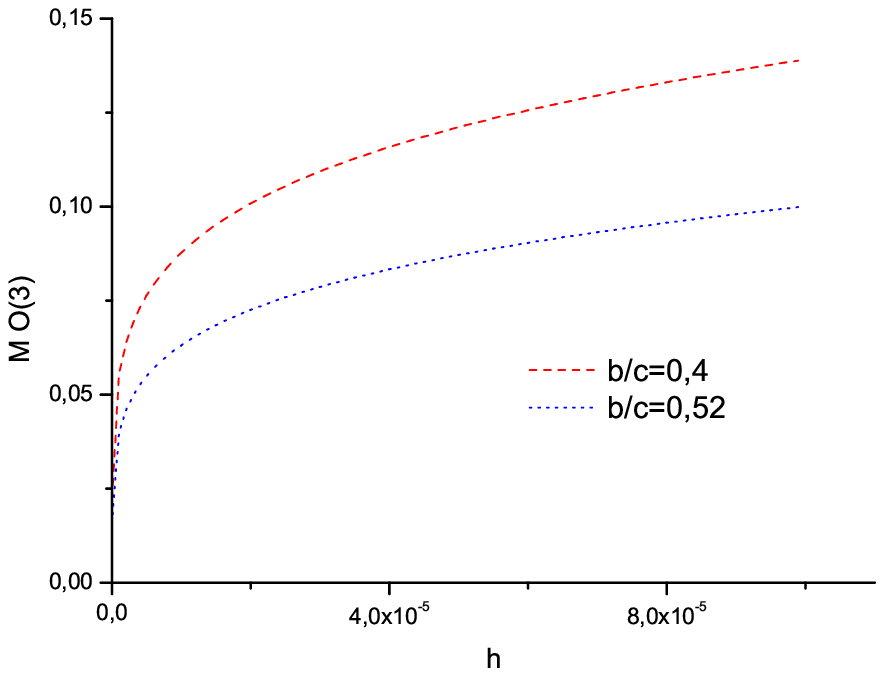}%
	\caption{The dependence of the order parameter on the field for $\tau=0$ and n=3}%
\label{M3}%
\end{figure}

\begin{figure}
    \centering
        \includegraphics[width=\figsize]{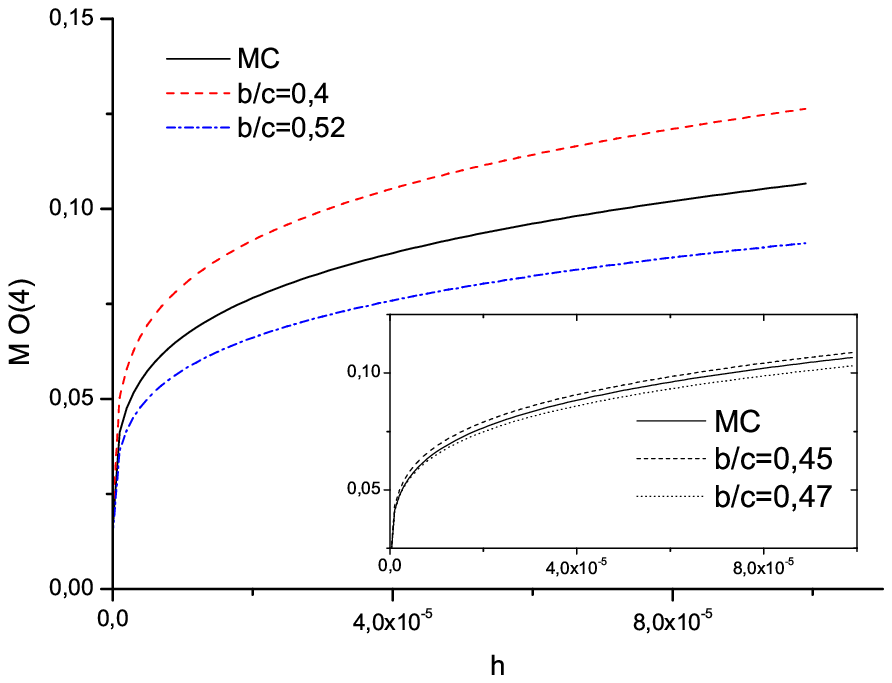}
    \caption{The dependence of the order parameter on the field for $\tau=0$ and n=4}
    \label{M4}
\end{figure}

Thus we obtained explicitly the equation of state of the n-vector
model in the presence of the external field. Its form allows easily
to pass to boundary cases of dependence only on the temperature or
on the field. So this equation is called the crossover equation. The
quantity $\sigma_{00}^+$ is the scaling function of the crossover
equation of state. It depends on the ratio of the field to the
temperature $\alpha$. Eq.(\ref{M}) allows to obtain a graph of
dependence of the order parameter on the field for $T=T_c$ and
compare it with results of Monte-Carlo simulations for analogous
models. Graphs of such dependencies for different n and different
parameters of the interaction potential are presented on
fig.\ref{M1}, \ref{M2}, \ref{M3}, \ref{M4}. As we can see from these figures the order parameter $M$ decrees when  the componence of the model n increases.

\begin{figure}
	\centering
  \includegraphics[width=\figsize]{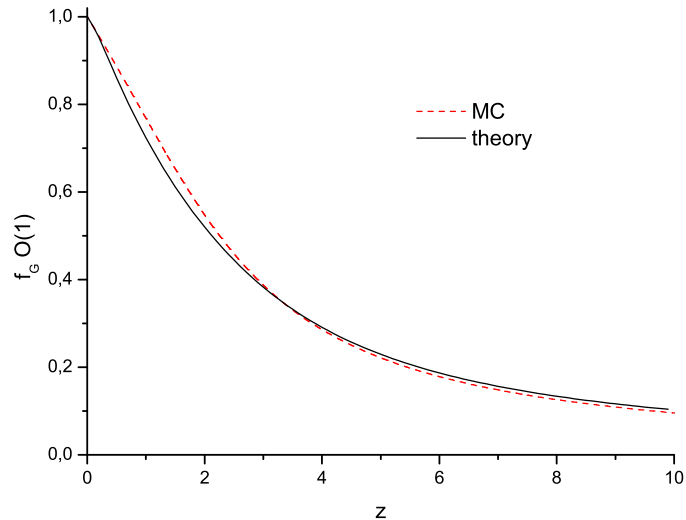}
  \caption{The scaling function for n=1 and $b/c=0.5$, solid curve -- our results, dashed curve -- Monte-Carlo data \cite{Eng1}.}\label{fig_f_G}
\end{figure}

\begin{figure}
    \centering
        \includegraphics[width=\figsize]{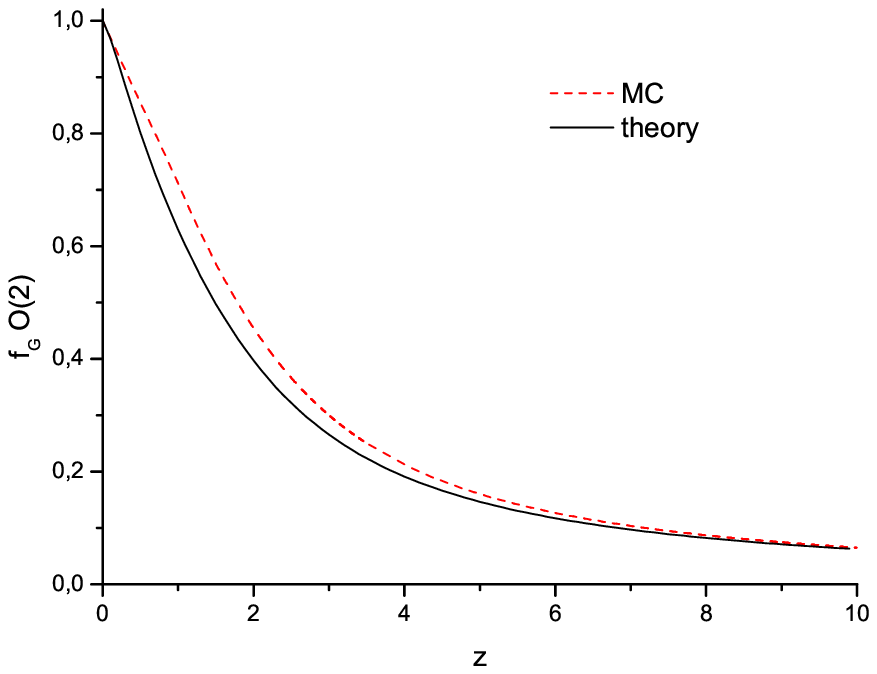}
    \caption{The scaling function for n=2 and $b/c=0.5$, solid curve -- our results, dashed curve -- Monte-Carlo data \cite{Eng2}.}
    \label{f_G2}
\end{figure}

\begin{figure}%
	\centering
		\includegraphics[width=\figsize]{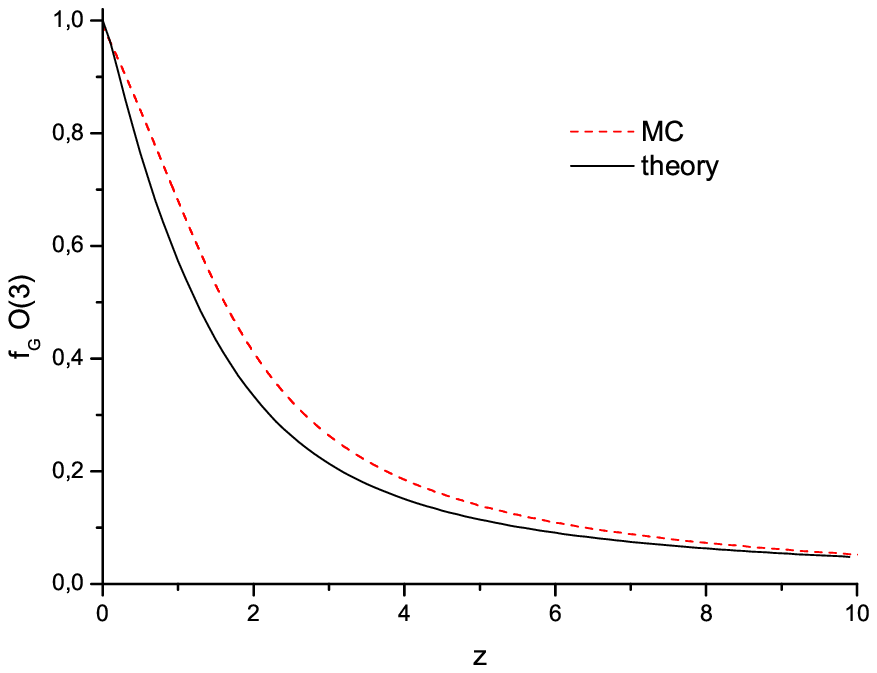}%
	\caption{The scaling function for n=3 and $b/c=0.5$, solid curve -- our results, dashed curve -- Monte-Carlo data 			\cite{Camp}.}%
	\label{f3}%
\end{figure}

\begin{figure}
	\centering
  	\includegraphics[width=\figsize]{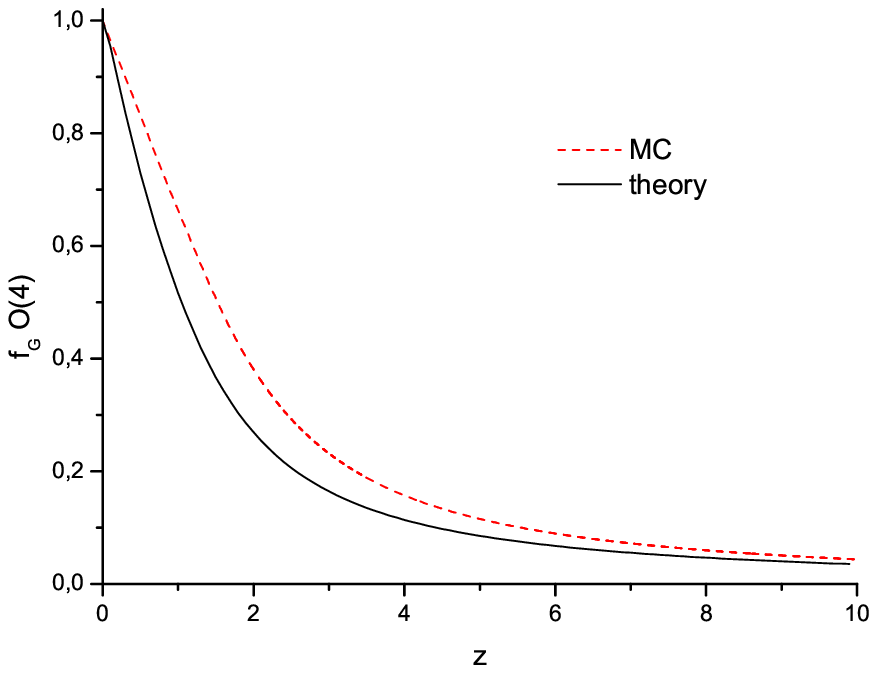}
  \caption{The scaling function for n=4 and $b/c=0.5$, solid curve -- our results, dashed curve -- Monte-Carlo data \cite{Eng4}.}\label{fig_fg4}
\end{figure}

There are different forms of the equation of state. Some discussion
about convenience the correspondent forms of the equation of state
is presented in \cite{21}. The equation of state (\ref{M}) can be reduced to
the form used in \cite{Eng1, Eng2, Eng4}: \be M=h^{1/\delta}f_G(z),
\ee where: \be h=H/H_0, \;z=\frac{\bar t}{h^{\beta\delta}},\; \bar
t=\tau\frac{T_c}{T_0}, \ee and $f_G$ is the scaling function. 
The explicit form of $f_G$ can be found from extrapolation of the MC data obtained  in \cite{Eng1, Eng2, Eng4, Camp}.
 $H_0$
and $T_0$ are normalization constants. Such form is equivalent to
the Widom-Griffiths equation of state \cite{Grif} \be y=f(x) \ee
where \be y\equiv h/M^\delta, \; x\equiv t/M^{1/\beta}. \ee The
scaling function obtained from Eq.(\ref{M}) has the form \be
f_G=\left(s_0^{3/2}/h_o\right)^{\frac{1}{5}}\sigma_{00}\left(1+\alpha^{-2}\right)^{\frac{1}{10}}.\label{f_G}
\ee It depends on $\alpha$ \be \alpha=\frac{\tilde h}{\tilde
\tau^{p_0}}. \ee The variables $\alpha$ and $z$ are connected
with the ratio: \be
\alpha=\frac{s_0^{3/2}}{h_o}\left(\frac{f_n}{c_{1k}}\right)^{p_0}z^{-p_0},
\ee that allows to compare our results with Monte-Carlo data.
Figures \ref{fig_f_G}-\ref{fig_fg4} present graphs of the
scaling functions for different n, where the dashed curve is the
Monte-Carlo data.

\section{Conclusions.}
We obtained the partition function of the n-vector model in the
presence of the external field above the critical temperature by the
CV method. The method of calculation corresponds to the general
scheme of the RG approach. Taking into account the explicit form of
the interaction potential allows to obtain the explicit dependence
of the coefficients of the linearized recurrence relations on the
temperature and the microscopic parameters of the model.

The explicit form of the exit point from the CR allows to obtain the
equations for the recurrence relations in the CR suitable for any
ratios of the temperature and the field. The structure of the
partition function as a product of the partial partition functions
that present different fluctuation processes allows to obtain the
explicit form for the free energy of the system. The order parameter
of the model was found by direct differentiation by the field.

The formulas to describe the field dependencies of the order
parameter of the n-vector model with exponentially decreasing
interaction potential for different ratios $b/c$ ($b$ is range of
the interaction potential, $c$ is period of the simple cubic
lattice) were obtained. It was found that for each value of n there
is an appropriate value of $b$ for which our dependencies are close
to Monte-Carlo data (see Fig.\ref{fig_f_G}-\ref{fig_fg4}).

The explicit form of the scaling function (\ref{f_G}) was found. The
comparison with Monte-Carlo data shows some difference for the
behavior of $f_G(z)$ for intermediate values of $z$. It may be
caused by used approximation in which the critical exponent $\eta=0$
and corrections to scaling were neglected. But further specification
of calculations is a subject of a separate investigation.

\section{Appendix}
\bea \varphi(x)&=&(n+2)U_n^2(x)+2xU_n(x)-2,\non
U_n(x)&=&\frac{U\left(\frac{n+1}{2},x\right)}{U\left(\frac{n-1}{2},x\right)},\;
x=\sqrt{\frac{3}{a_4}}d(B_1,B'),\no \eea $U(a,x)$ is the function of
Weber's parabolic cylinder. \bea
Q(d_l)&=&(2\pi)^{\frac{n}{2}}\left(\frac{3}{a_4^{(l)}}\right)^{\frac{n}{4}}U\left(\frac{n-1}{2},x_l\right)\exp\left({\frac{x_l^2}{4}}\right),\non
a_4&=&-3s_0^d\frac{n^2}{m^4}\left(1-z'U_n(z')-U_o^2\right),\non
U_0&=&\sqrt{\frac{n+2}{2}}U_n(z'),\;
z'=\sqrt{\frac{3}{u'_4}}u'_2,\non
r_{p+m}&=&\frac{U'_n(x_{n_p+m})}{U_n(x_{n_p+m})}-\frac{1}{2}\frac{\varphi'(x_{n_p+m})}{\varphi(x_{n_p+m})},\non
g_{p+m}&=&-\frac{\bar x E_2^{m-1}H_{cd}}{\sqrt{1+\Phi
E_2^{m-1}H_c}}\left(1-\frac{H_c\Phi E_2^{m-1}}{2\left[1+\Phi
E_2^{m-1}H_c\right]}\right),\non \bar
x&=&f_n\sqrt{\frac{3}{\phi_n}}\no \eea


\begin{thebibliography}{0}
\expandafter\ifx\csname natexlab\endcsname\relax\def\natexlab#1{#1}\fi
\expandafter\ifx\csname bibnamefont\endcsname\relax
  \def\bibnamefont#1{#1}\fi
\expandafter\ifx\csname bibfnamefont\endcsname\relax
  \def\bibfnamefont#1{#1}\fi
\expandafter\ifx\csname citenamefont\endcsname\relax
  \def\citenamefont#1{#1}\fi
\expandafter\ifx\csname url\endcsname\relax
  \def\url#1{\texttt{#1}}\fi
\expandafter\ifx\csname urlprefix\endcsname\relax\def\urlprefix{URL }\fi
\providecommand{\bibinfo}[2]{#2}
\providecommand{\eprint}[2][]{\url{#2}}

\end{thebibliography}


\begin{thebibliography}{10}

\bibitem{Bom}
D.~Bohm,
\newblock {\em General Theory of Collective Variables} (Mir, Moscow, 1964) (in Russian).

\bibitem{Zub}
D.~N. Zubarev,
\newblock Dokl. Akad. Nauk SSSR {\bf 95}, 757 (1964) (in Russian).

\bibitem{06}
I.~R. Yukhnoskii,
\newblock {\em Phase Transitions of the Second Order. Collective Variables
  Method} (World Scientific, Singapore, 1987).

\bibitem{Wil}
K.~G. Wilson,
\newblock Phys. Rev. B {\bf 4}, 3174 (1971).

\bibitem{Pok}
A.~Z. Patashinskii and V.~L. Pokrovskii,
\newblock Sov. Phys. JETP {\bf 23}, 292 (1966).

\bibitem{Kad}
L.~P. Kadanoff,
\newblock Physics {\bf 2}, 263 (1966).

\bibitem{Avd}
G.~M. Avdeiva and A.~A. Migdal,
\newblock Sov. Phys JETP Lett. {\bf 16}, 178 (1972) (in Russian).

\bibitem{Briz}
E.~Bresin, D.~J. Wallace, and K.~G. Wilson,
\newblock Phys. Rev. Lett. {\bf 29}, 591 (1972).

\bibitem{Briz1}
E.~Bresin, D.~J. Wallace, and K.~G. Wilson,
\newblock Phys. Rev. B {\bf 7}, 232 (1973).

\bibitem{Weg}
F.~J. Wegner,
\newblock Phys. Rev. B {\bf 5}, 4529 (1972).

\bibitem{Weg1}
E.~K. Riedel and F.~J. Wegner,
\newblock Phys. Rev. B {\bf 9}, 1238 (1973).

\bibitem{Fish}
M.~E. Fisher and A.~Aharony,
\newblock Phys. Rev. B {\bf 10}, 2818 (1974).

\bibitem{Nik}
J.~F. Nicoll and P.~C. Albright,
\newblock Phys. Rev. B {\bf 31}, 4576 (1984).

\bibitem{Nel}
D.~R. Nelson,
\newblock Phys. Rev. B {\bf 11}, 3504 (1974).

\bibitem{Bag}
C.~Bagnuls and C.~Bervillier,
\newblock J. Phys. Lett. (Fr.) {\bf 45}, L95 (1984).

\bibitem{Domb1}
V.~Dohm,
\newblock Z. Phys. B-Condensed Matter {\bf 60}, 61 (1956).

\bibitem{Domb2}
H.~J. Krause, R.~Schloms, and V.~Dohm,
\newblock Z. Phys. B-Condensed Matter {\bf 79}, 287 (1990).

\bibitem{Ok1}
Y.~Okabe and K.~Ohno,
\newblock J. Phys. Soc. Jpn {\bf 53}, 3070 (1984).

\bibitem{Ok2}
Y.~Okabe and K.~Ideura,
\newblock Prog. Theor. Phys. {\bf 66}, 1959 (1981).

\bibitem{Brez}
E.~Bresin, J.~C.~L. Guillon, and J.~Zinn-Justin,
\newblock Phys. Lett. A {\bf 47}, 285 (1974).

\bibitem{Fer1}
A.~M. Ferrenberg and D.~Landau,
\newblock Phys. Rev. B {\bf 44}, 5081 (1991).

\bibitem{Fer2}
K.~Chen, A.~M. Ferrenberg, and D.~P. Landau,
\newblock Phys. Rev. B {\bf 48} (1993).

\bibitem{Eng1}
J.~Engels, L.~Fromme, and M.~Seniuch,
\newblock Nucl. Phys. B {\bf 655}, 277 (2003).

\bibitem{Eng2}
J.~Engels, S.~Holtmann, T.~Mendes, and T.~Schulze,
\newblock Phys. Lett. B {\bf 492}, 219 (2000).

\bibitem{Eng4}
J.~Engels and T.~Mendes,
\newblock Nucl. Phys. B {\bf 572}, 289 (2000).

\bibitem{Camp}
M.~Campostrini, M.~Hasenbusch, A.~Pelissetto, P.~Rossi, and E.~Vicari,
\newblock Phys. Rev. B {\bf 65}, 40 (2002).

\bibitem{VakRudGol}
I.~A. Vakarchuk, Y.~K. Rudavskii, and Y.~V. Holovach,
\newblock Physics of Many Particle Systems {\bf 4}, 44 (1983) (in Russian) .

\bibitem{YuhMr2}
I.~R. Yukhnoskii,
\newblock {\em Selected works. Physics} (Lviv Polytechnic National University,
  Lviv, 2005) (in Ukrainian).

\bibitem{Usat}
Z.~E. Usatenko and M.~P. Kozlovskii,
\newblock Phys. Rev. B {\bf 62}, 3599 (2000).

\bibitem{Usat1}
Z.~E. Usatenko and M.~P. Kozlovskii,
\newblock Mater. Sci. Eng., A {\bf 227}, 732 (1997).

\bibitem{Juh}
I.~R. Yukhnovskii, M.~P. Kozlovslii, and I.~V. Pylyuk,
\newblock {\em Microscopic theory of phase transitions in the three-dimensional
  systems} (Eurosvit, Lviv, 2001) (in Ukrainian).

\bibitem{21}
M.~P. Kozlovskii,
\newblock Ukr. Fiz. Zh./Reviews (Ukr. ed.) {\bf 5}, 61 (2009).

\bibitem{20}
M.~P. Kozlovskii, I.~V. Pylyuk, and O.~O. Prytula,
\newblock Cond. Matt. Phys. {\bf 7}, 361 (2004).

\bibitem{Prytula}
M.~P. Kozlovskii, I.~V. Pylyuk, and O.~O. Prytula,
\newblock Nucl. Phys. B {\bf 753}, 242 (2006).

\bibitem{Grif}
R.~B. Griffiths,
\newblock Phys. Rev. {\bf 158}, 176 (1967).

\end{thebibliography}
\end{document}